\documentclass[conference]{IEEEtran}
	\IEEEoverridecommandlockouts
    \usepackage{fancyhdr,lipsum}

    \fancypagestyle{mahmood}{%
   \fancyhf{} 
   
   \fancyhead[C]{IEEE Black Sea Conference on Communication and Networking, June 2022, Sofia Bulgaria.}
    }%

    \makeatletter
    \let\ps@IEEEtitlepagestyle\ps@mahmood
    \makeatother
	\usepackage{cite}
	\usepackage{amsmath,amssymb,amsfonts}
	\usepackage[noend]{algorithmic}
	\usepackage[dvisvgm, usenames, dvipsnames]{color}
	\usepackage{graphicx}
	\usepackage{textcomp}
	\usepackage{xcolor}
	\usepackage{verbatim}
	\usepackage{afterpage}
	\usepackage{ragged2e}
	\usepackage{comment}

    \def\BibTeX{{\rm B\kern-.05em{\sc i\kern-.025em b}\kern-.08em T\kern-.1667em\lower.7ex\hbox{E}\kern-.125emX}}
    
	\usepackage[margin=1in,footskip=0.25in]{geometry}
\begin{document}
	
		\title{Ransomware Detection and Classification Strategies}
	\author{\IEEEauthorblockN{Aldin Vehabovic$^{1}$, Nasir Ghani${^1}$, Elias Bou-Harb$^{2}$, Jorge Crichigno$^{3}$, Ayseg$\ddot{\mbox{u}}$l Yayimli${^4}$ \\
	\textit{$^{1}$Univ. of South Florida, $^{2}$Univ. of Texas San Antonio, $^{3}$Univ. of South Carolina, $^{4}$Valparaiso University}}
	}
	\maketitle

\maketitle

\begin{abstract}
Ransomware uses encryption methods to make data inaccessible to legitimate users. To date a wide range of ransomware families have been developed and deployed, causing immense damage to governments, corporations, and private users. As these cyberthreats multiply, researchers have proposed a range of ransomware detection and classification schemes. Most of these methods use advanced machine learning techniques to process and analyze real-world ransomware binaries and action sequences. Hence this paper presents a survey of this critical space and classifies existing solutions into several categories, i.e., including network-based, host-based, forensic characterization, and authorship attribution. Key facilities and tools for ransomware analysis are also presented along with open challenges.
\end{abstract}
%
	\emph{Cybersecurity, ransomware, machine learning}

\section{Introduction}
Ransomware is a form of malware that encrypts user files on a host computer or server. Once these operations have been completed, malicious actors (malactors) demand some form of ransom payment (monetary or otherwise) from their victims to release the decryption keys.  However, in many cases victims may still not get their data back even after paying the ransom. Hence ransomware has emerged as a very serious threat and is already the most profitable type of malware (with total annual payments in the billions of dollars) \cite{cisco2016}. This spread has been further bolstered by cryptocurrencies offering high anonymity and complicating the tracking/identification of attackers.

Now the earliest example of ransomware emerged over 3 decades ago when compromised \textit{compact disks} (CD) were mailed to conference attendees.  Subsequently, ransomware evolved to more network-based delivery  (Archiveus Trojan and Gpcodoe, 2006) and cryptocurrency payments (WinLock, 2008).  Today a wide range of families have been deployed, impacting many users and organizations, e.g., government agencies, utility providers, healthcare organizations, manufacturing and technology firms, etc. In particular, there has been an uptick in ransomware attacks against organizations since associated payoffs are generally much larger \cite{trends2022}. By some accounts, up to 50\% of large organizations experienced ransomware attacks in 2020 \cite{kapoor2022}. Among many notable incidents, the 2021 Colonial Pipeline ransomware attack caused sizable disruption to fuel supplies in the US Northeast. In the same year, a large computer manufacturer, Acer Inc., also had its financial data impacted by a ransomware attack, and the Irish healthcare system was also targeted.

Ransomware continues to evolve with increasing levels of secrecy and sophistication, posing an unacceptable level of risk to governments, corporations and private users. Therefore many efforts are being made to address the challenges in this domain. For example, the US government is proposing legislation to mandate reporting of ransomware payments, and some are even proposing an outright ban on ransom payments \cite{law2021}.  Similar directives are also being considered in the European Union (EU). Meanwhile, researchers are actively developing new ransomware detection and mitigation strategies.  The overall approach here is to use ransomware signatures to detect nefarious activities and update network and host defences to prevent or limit their operation, e.g., in firewalls and host-based antivirus programs.  It is here that \textit{machine learning} (ML) methods, including \textit{neural network} (NN) based schemes, have proven to be very effective in analyzing large amounts of empirical data and training advanced ransomware detection and classification algorithms. 

Hence this paper presents a survey of some key contributions in ransomware detection and classification.  Since early detection is the most effective solution, further recovery methods are not reviewed. This work differs from earlier surveys \cite{berrueta2019},\cite{mouss2021} as it presents a broader taxonomy and also reviews analysis tools. First, Section \ref{section_attacks} details the ransomware ``kill chain'' and highlights other aspects.  Section \ref{section_network} then overviews network-based detection, whereas Section \ref{section_host} focuses on localized host-based detection. Further methods using forensic analysis and characterization are also reviewed in Section \ref{section_forensic} along with malware authorship attribution in Section \ref{section_attribution}. Key facilities and tools for ransomware analysis are then detailed in Section \ref{tools} followed by open challenges in Section \ref{challenges}.  Also note that the focus here is on Windows-based ransomware as this is the most common \textit{operating system} (OS) today.
\begin{figure*}[ht]
    \centering
    \includegraphics[width=6.5in, height=2.15in]{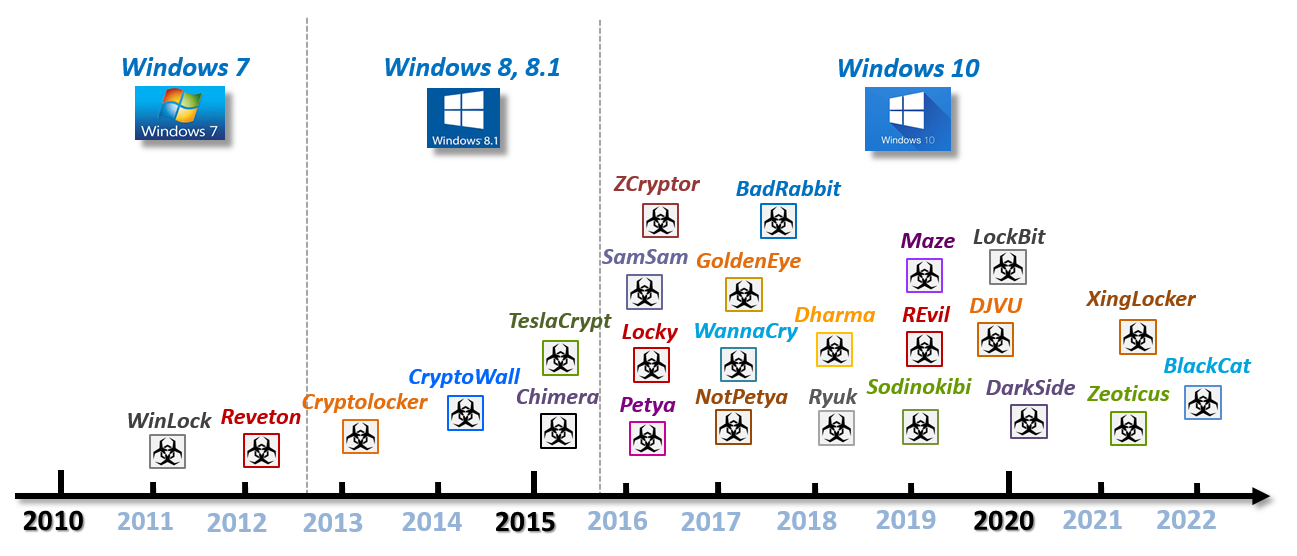}
    \caption{Timeline of ransomware families (Windows-based)}
    \label{ransomware_timeline}
\end{figure*}

\section{Overview of Ransomware}
\label{section_attacks}
The ransomware ecosystem consists of multiple actors. Most notably, this includes targeted user hosts and malicious \textit{command and control} (C\&C) servers operated by malactors (to handle encryption keys and payments). Other intermediaries also play important supporting roles here, e.g., to identify victims, distribute binaries, process payments, etc. Akin to other malware services, new \textit{ransomware as a service} (RAAS) offerings have also emerged, allowing malicious users to directly purchase attacks from ``RaaS affiliates''.  Indeed, this greatly lowers the entry barrier for conducting such cyberattacks.

\begin{figure*}[ht]
    \centering
    \includegraphics[width=5.45in, height=2.45in]{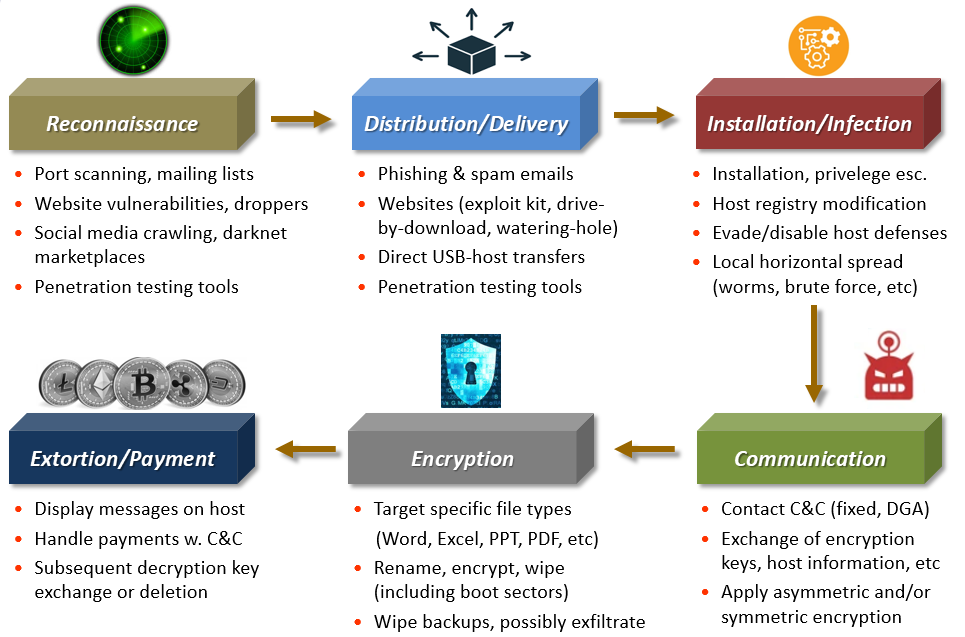}
    \caption{Overview of ransomware attack ``kill chain''}
    \label{ransomware}
\end{figure*}

Overall, there are several types of ransomware, and a timeline of some major families is shown in Figure \ref{ransomware_timeline}. For example, locker ransomware is designed to lock users out of their machines and demand payments. Meanwhile, cryptographic ransomware encrypts user files and demands ransoms, and this type is the most common. Finally, double-extortion ransomware threatens further data release (also called doxing).  Now ransomware operation entails a sequence of stages, termed as the ``kill-chain'', Figure \ref{ransomware}.  Although others have shown slightly different versions of this sequence with varied stages and namings, the key operations are still the same.  These stages are now briefly detailed, see also \cite{berrueta2019}:
\begin{itemize}
    \item \underline{\textbf{Reconnaissance}}: This initial stage focuses on identifying (enumerating) a list of potential hosts to target for ransomware transmission. Hackers or RaaS affiliates can use a range of methods here, including port scanning, mailing lists, Internet/social media crawling, or directly purchasing lists from darknet marketplaces, etc. 
    \item \underline{\textbf{Distribution/Delivery}}: The next step focuses on delivering ransomware binaries to the identified hosts. Again, a wide range of techniques are used here, e.g., spam/phishing emails, website exploits (drive-by attacks), even manual transfers using removable drives.  As expected, there is almost always an (inadvertent) human element involved in downloading malware onto a device.
    \item \underline{\textbf{Installation/Infection}}: This stage entails ransomware setup on infected hosts. Most advanced strains also try to hide their entry/presence by doing various things, e.g., limiting pre-attack ``paranoia'' activities, uncovering/disabling backups, blocking host defense/firewalls, etc.  Spreading (propagating) ransomware designs also perform internal reconnaissance to identify other hosts to infect, i.e., worm-like operation.
    \item \underline{\textbf{Communication}}: This stage usually runs prior to encryption and involves communicating with the C\&C server. The details here can vary based on the type of encryption being used. For example, symmetric encryption designs generate a local key which is either sent to the external C\&C server or stored locally (but encrypted with the attacker's public key).  Hence victims must contact the C\&C server to obtain the decryption keys. However, symmetric encryption is vulnerable to interception by anti-virus programs as it stores decryption keys on local hosts (at least for some time). Hence other standalone designs use asymmetric public key encryption along with the attacker's public keys.  However, such encryption is generally slower and more vulnerable to detection by host anti-virus programs.
    \item \underline{\textbf{Encryption}}: This is the main step where ransomware runs encryption algorithms to lock user data and/or machine access.  The original data is usually wiped (along with any detected backups) and a message of some sort is displayed.  However, excessive calls to encryption routines can take time and also consume a lot of processor cycles.  In turn, these signatures can be detected by host defenses. Hence some ransomware strains try to maximize their impact by only encrypted a small portion of a file (but still enough to render it useless to users).
    \item \underline{\textbf{Extortion/Payment}}: This final stage involves the actual handling of ransom payments and any terminal action sequences.  Again, many ransomware designs request payments in cryptocurrencies or through the darkweb.  Depending on the intentions of the malactors, some ransomware designs may not even release encryption keys after payment.
\end{itemize}
\begin{table*}
	\centering
\caption{Common Windows-based cryptographic ransomware families}
\label{table_ransomware}
\begin{tabular}{|l|c|l|} 
\hline
\textbf{Name} & \textbf{Origin} & \textbf{Characteristics} \\ \hline \hline
Reveton & 2012 & Screen locker type (with variants also doing file encryption), distribution via malicious websites \\ \hline
Cryptolocker & 2013 & Website and email distribution, AES-256 symmetric encryption, C\&C messaging via onion routing  \\ \hline
CryptoWall & 2014 & Copied into registry keys/startup folders, 2048 bit RSA asymmetric encryption, high revenues generated  \\ \hline
TeslaCrypt & 2015  & Trojan ransomware, strong distribution networks, attacks gaming files, encrypts small files (symmetric encryption) \\ \hline
Chimera & 2015 & Phishing distribution, encrypts local and network drives  using symmetric encryption,  also threatens release (doxing) \\ \hline
SamSam & 2016 & Non-spreading, RDP access w. privilege escalation, RSA-2048 asymmetric encryption \\ \hline
Locky & 2016 &  Email delivery with malicious MS Word (macros), encrypts certain file types, RSA-2048 and AES-128 encryption \\ \hline
Petya & 2016 & Cloud-based distribution, self-propagating (worm), admin privileges, encrypts boot table (NTFS) and data files \\ \hline
GoldenEye & 2016  & Like Petya but target files (.exe), malicious Excel file (macros), AES asymmetric encryption, offered through RaaS \\ \hline
Zcryptor & 2016  & Worm-like (spreading) via spam (files and macros), encrypts local/ shared drives and USB (multiple file formats) \\ \hline
NotPetya & 2017 &  Worm-like (spreading), exploit Windows SMB protocol, AES-128 encryption, target nations and global shipping \\ \hline
WannaCry & 2017 & First “global” ransomware, fast spreading, 2048-bit RSA encryption, targeted governments, education, etc  \\ \hline
REvil & 2019 & First to target IT providers, nation-state origin, stream cipher encryption (RC-4), threaten publication (doxing) \\ \hline
Bad Rabbit & 2020 & Drive-by attack distribution, spreading type, targets users in specific regions, AES-128 and RSA-2048 encryption \\ \hline
DarkSide & 2020 & Cybercriminal gang (RaaS), spreads across networks, RSA encryption, targets large corporations, data exfiltration \\ \hline
Zeoticus & 2021 & Run offline, avoids certain regions, changes registry keys (persistent), hybrid encryption, Proton mail payments\\ \hline
\end{tabular}  
\end{table*}
Note that large files can lead to lengthy encryption times and high processor usage. In turn, this can increase the chances of detection/mitigation by host anti-virus programs \cite{berrueta2019}.  As a result, some ransomware designs only encrypt parts of a file to render it unusable. Others may leave the first few bytes of a file (metadata) unencrypted to complicate tampering detection. Overall, ransomware differs from other forms of malware in several important ways.  For example, impacted hosts are not necessarily controlled en-masse by botnet controllers, unlike in \textit{distributed denial-of-service} (DDoS) attacks.  Instead, ransomware C\&C servers operate much more intermittently and discretely (i.e., responding to communication requests, handling keys, and processing payments).  Furthermore, unlike worm malware, not all ransomware is spreading/propagating as this increases the likelihood of detection. Table \ref{table_ransomware} summarizes some of the key ransomware families that have emerged over the last decade.  A range of associated detection and classification strategies are now reviewed here.

\section{Network-Based Detection Schemes}
\label{section_network}
Network-based detection schemes analyze host traffic for ransomware activities, i.e., C\&C communications (Figure \ref{ransomware}). Here, packet data can be collected from infected hosts or enterprise/carrier networks. Overall, several types of network traces can indicate ransomware activity.  For example, many strains send \textit{domain name service} (DNS) queries to resolve C\&C server IP addresses, i.e., either statically hardcoded in binaries or generated in a psuedo-random manner. Note that these DNS requests can be detected/blocked by anti-virus programs by extracting static names and/or fingerprinting dynamically generated ones. Other traffic types can also be analyzed, e.g., networked storage access, etc. Some contributions in this space are now reviewed.

The authors in \cite{almash2019} presents a network detection system for Locky ransomware. A testbed is used to run multiple samples of this malware, and then behavioral and non-behavioral traffic features are analyzed, e.g., HTTP-POSTS, MDN, and DNS (IPv4, IPv6). A multi-classifier \textit{intrusion detection system} (IDS) is then developed using these features to detect packet- and flow-level behaviors using a range of ML algorithms, i.e., \textit{random forest} (RF), random tree, Bayes network, \textit{support vector machines} (SVM), etc. Results show a mean detection rate of 97\%.  Meanwhile, \cite{mulders2017} states that most detection methods are very labor intensive, e.g., requiring sandbox runs, reverse engineering (original code), or host-based countermeasures. Hence a more automated networking-based approach is presented for the Samba protocol, a \textit{server message block} (SMB) protocol for shared network files. A ransomware detection scheme is then outlined to detect and interfere with incoming attacks (by disabling connections on an infected machine). Various methods are implemented here, including analyzing ransomware in real-time, studying the characteristics of log data, and using behavioral analysis to distinguish between normal and malicious traffic. Another SMB-focused solution called REDFISH is also outlined in \cite{morato2018}, and this scheme analyzes transfers using a network probe. A heuristic time-series method is then used to detect ransomware activity and block file encryption.  Results show a 100\% detection rate for 19 different ransomware families.  

Furthermore, \cite{cusak2018} presents a ML ransomware detection system using a \textit{software defined network} (SDN) framework. The goal here is to detect an attack before encryption by using \textit{programmable forwarding engines} (PFE) to monitor packets. This data is analyzed to detect communications between infected hosts and C\&C servers, and detection is done using a RF binary classifier. Results show a detection rate of 86\% and a false negative rate of 11\%.  Additionally, \cite{lokuketagoda2018} presents its R-Killer scheme to target the distribution stage by identifying and blocking ransomware emails.  The solution has three sub-systems, i.e., core detection engine (for emails), a sandbox environment, and a proactive monitoring entity (website links). Note that the key difference between R-Killer and email scanning tools is that the former directly communicates with email servers to retrieve selected emails for scanning purposes. The system also builds a threat intelligence repository without leaking user data.

Meanwhile, the work in \cite{homayoun2019} details a rapid ransomeware detection and identification scheme (within 10 seconds of execution).  This approach uses two NN designs to classify host activities, i.e., \textit{long short-term memory} (LSTM) and \textit{convolutional neural networks} (CNN). Four ransomware strains are considered, and training is done using 220 samples of each.  The model achieves a true positive rate of 97.2\%, and the system can identify previously undetectable ransomware, i.e., CryptoWall, Torrent locker, and Sage. As such, this solution offers some promise for zero-day threat detection as well. Meanwhile \cite{kurniawan2018} presents network forensics behavior analysis for the Cerber ransomware (which offloads target host search and encrypts all files). The solution analyzes packet headers, protocol types, and payloads (for viruses and spam). Signature detection is then used to find similarities with previous attacks. Furthermore, \cite{su2019} studies ransomware in Chinese social networks and highlights problems such as locking screens, call blocking, hijacking, and password compromises. Feature extraction and ML are also used to detect locker ransomware by analyzing messaging transactions, e.g., via SVM, RF, and logistic regression.  Results show a detection accuracy of 99\%. Meanwhile, \cite{kao2018} focuses on WannaCry and uses the \textit{Sysinternal} and \textit{Wireshark} tools for behavioral and network analysis to identify compromise indicators. Results confirm successful tracking of WannaCry based on registry modification, processes, and file system and network activity.

Also, \cite{roy2021} presents a deep learning scheme for ransomware detection/classification called DeepRan. Here, hosts use an attention-based \textit{bi-directional LSTM} (Bi-LSTM) network scheme to analyze network activity and classify it as either normal or abnormal. Up to 17 families are studied, and a testbed is used to collect logs from two users over 63 days.  Results indicate a detection accuracy of 99.87\% and successful prevention of ransomware infections. Also, \cite{krzysztof2018} introduces a SDN-based approach for ransomware detection that examines communications for CryptoWall and Locky ransomware (website messages, content sizes, etc). Three key components are implemented  here, including detection method, learning phase, and fine-tuning phase. Findings show a detection rate of 97-98\% with little damage to infected hosts. NetConverse also uses ML-based schemes for network detection \cite{alhawi2018}. This scheme analyzes Windows-based host traffic and implements three phases, i.e., data collection, feature extraction, and ML classification. In particular, a range of algorithms are used for the latter phase, including Bayesian networks, decision trees, \textit{k-nearest neighbour} (k-NN), multi-layer perceptron, RF, etc. Results show a detection rate of 97.1\% for decision trees and 99.9\% for RF classifiers.

\section{Host-Based Detection Schemes}
\label{section_host}
Host-based detection schemes monitor local system activities to detect malware, both before and after attacks.  These capabilities are typically bundled into antivirus programs, and a range of static and dynamic actions (and frequencies) are tracked, e.g., such as memory and file operations, \textit{application programmer interface} (API) function calls, \textit{dynamic link library} (DLL) calls, etc.  For example, file operations can include deletion, overwriting/modification, extension changes, directory accesses, etc. Some key host-based schemes for ransomware detection are now presented.

The work in \cite{kharraz2016} presents a dynamic analysis solution for ransomware detection called UNVEIL. This scheme uses the Cuckoo sandbox to monitor system and file actions, e.g., persistent desktop messages (API display message calls), selective encryption/deletion of files based on attributes (size, date, accessed, and extension), etc. The design requires access to file system modification information and analyzes data buffers in \textit{input/output} (I/O) requests. Tests are done using a dataset with 148,223 malware samples, and findings show that UNVEIL can successfully detect 13,637 ransomware samples with a true positive detection rate of 96.3\%. Meanwhile, \cite{kolodenker2017} presents a PAYBREAK scheme for hybrid encryption ransomware using symmetric encryption. Here, the encryption keys are monitored/stored in real-time, thereby facilitating file decryption (i.e., recovery). The proposed setup has three components, including cryptographic function hooking, key vault, and file recovery.  Results confirm that PAYBREAK can mitigate attacks from 12 out of 20 ransomware families (9 of which are unmitigated beforehand).

Also, the work in \cite{continella2016} proposes an add-on driver to immunize Windows-based file systems from ransomware attacks. Termed as SHEILDFS, this system uses an adaptive model to profile behaviors by analyzing billions of low-level file I/O operations for benign applications. Various parameters are tracked such as the number of folder listings, number of files read/written/renamed, file types, etc. The system is monitored and any operations are proactively rolled back if malicious activity is detected. Similarly, \cite{kharraz2017} analyzes application I/O requests to scan for ransomware activity and flag/restore affected files. These files are also changed to protected status to prevent any modifications before data is sent to the kernel. The model is tested using 504 samples from 12 families and evaluated using 9,432 samples. Results show a total of 1,174 samples being flagged as active ransomware strains from 29 families.  Meanwhile, \cite{baek2017} observes block request headers and classifies ransomware into 3 types based on how it overwrites encrypted files, i.e., Class A (in-place), Class B (out-of-place), and Class C (deleting and overwriting original). A detection algorithm is then presented to monitor 4 I/O request parameters. The solution also supports an instant recovery mode that keeps a log of what files have been changed and stores the originals. Hence if the algorithm suspects a ransomware threat, it notifies the user for confirmation, and if so, changes the file to read-only. A two-step unsupervised ransomware detection system, RAPPER, is also proposed in \cite{sinha2018}. Namely, the first step monitors process and system activity to flag suspicious behaviors. Meanwhile, the second step analyzes this activity to generate a detailed assessment using anomaly detection to track \textit{hardware performance counters} (HPCs). Rather then modeling ransomware types, the work focuses on normal behaviors. Specifically, \textit{long short term memory} (LSTM) encoding is used for unsupervised detection using a time series approach. The authors also present extensions for file backup/recovery. 

Recent efforts have also analyzed \textit{early-state} ransomware ``paranoia'' activities. These ``pre-attack'' actions are used to detect environments and avoid fingerprinting and detection (by execution in virtual environments).  Namely, the working assumption here is that benign applications will not try to sense/detect much in their run-time environments.  For example, \cite{molina2021} proposes a framework to monitor pre-attack API calls. Dynamic malware analysis is done by executing ransomware samples (from \textit{VirusTotal} and \textit{VirusShare}) in the \textit{Cuckoo} sandbox to fingerprint API activities.  Next, \textit{natural language processing} (NLP) (word-based) methods are used to represent these activities and extract a compact number of behavioral features to improve ML scalability. A range of classifiers are then trained using the extracted features, including Naive Bayes, k-NN, RF, NN, LSTM, and Bi-LSTM. This effort builds and analyzes one of the largest ransomware datasets, and results confirm that many families do generate distinguishable API fingerprints.  In particular, results show very high detection rates, with RF classifiers averaging 95\%.  Meanwhile \cite{alsabeh2020} also presents another scheme to intercept API calls and detect ransomware paranoia activities in Windows-based environments. First, the authors build a training dataset using the \textit{VirusTotal} repository and extract 117 ransomware samples (across 30 contemporary families) along with related metadata (as provided by anti-malware vendors such as BitDefender, McAfee, and Kaspersky).  These samples are then run in user mode on a host and targeted/selected API calls are routed to a detour function for logging purposes. To prevent false positives, API function call data is also collected for 98 benign applications. Subsequently, all selected API functions are assigned a rank from 1-10 based on how many times they are called. If the score exceeds a carefully-chosen threshold, this scheme flags the associated application as ransomware and also takes mitigation actions to terminate it. Results show an accuracy of 91\% on training data and 84\% on testing data, albeit the latter also gives a higher false negative rate of about 22\%.  This is attributed to the fact that some ransomware types do not perform any pre-execution inspection activities (making them hard to detect via such methods). The authors also provide GitHub access to their code and datasets. Meanwhile, \cite{molina2022} presents a host-based \textit{ransomware prevention and mitigation} (RPM) framework using proactive API call monitoring.  Ransomware samples are run offline in a sandbox to fingerprint API function calls (i.e., Reveton, Locky, Teslacrypt, etc). Extracted features are then used to build frequency pattern trees and then applied in real-time to compare executing API sequences (and detect potential paranoia activities).

Carefully note that Windows 10 also supports ransomware mitigation via its built-in blocking feature that allows users to enable ``Controlled Folder'' access \cite{kapoor2022}. Specifically, this toggle only allows trusted applications to access these folders, and users can also add other applications to the trusted list. Although this capability is very effective in blocking many ransomware families, it is not enabled by default and requires user awareness and vigilance. Moreover, future ransomware desgins may try to reset/disable this setting. Finally, some hardware-based solutions have also emerged. For example, \cite{huang2017} details a \textit{solid state drive} (SSD) called FlashGuard which implements a firmware recovery system to support rapid recovery of encrypted data (using out-of-place writes). As such no backup is required here. FlashGuard is implemented on a 1 terabyte SSD drive which is programmed for basic read, write, and erase commands (and uses 15\% capacity over-provisioning for its operation).

\section{Forensic Analysis \& Characterization}
\label{section_forensic}
Forensic analysis focuses on recovering, gathering, and analyzing information from infected machines to determine the effects of malware (and uncover any identifying information). Accordingly, various studies have also applied these methods to ransomware. Consider the details.

The authors in \cite{poudyal2018} introduce a multi-level ML analysis framework to detect/classify ransomware. This scheme analyzes raw binaries, assembly code, and libraries using tools such as Linux object-code dump and a portable executable parser. ML classifiers are then trained using the extracted data, and results show a detection rate of about 90\% with various algorithms.  Meanwhile, \cite{zhang2019} presents one of the first schemes to classify/detect ransomware using static analysis. Here, operational code sequences are first transformed into N-grams. Next, the \textit{term frequency-inverse document frequency} (TF-IDF) is computed for each N-gram for classification purposes, and several classifiers are trained (including decision tree, RF, k-NN, naive Bayes). Experimental results show an impressive detection rate of about 91\%.

Note that many ransomware mitigation techniques compare run-time models or model code snippet semantics.  However, related accuracy can be low here, and hence \cite{ming2017} proposes a novel hybrid scheme called ``sliced segment equivalence checking'' to identify fine-grained semantic similarities/differences between executables. A prototype BinSim system is also built to successfully identify fine-grained relations between obfuscated binaries (and this outperforms existing binary diffing tools). Meanwhile, \cite{chen2017} proposes a dynamic ransomware detection system that uses various ML algorithms (such as RF, SVM, simple logistic, and naive Bayes) to identify known and unknown ransomware types. In particular, raw API data is used to build call flow graphs (features), and findings show detection rates of over 91\%.

Furthermore, \cite{quinkert2018} introduces its RAPTOR scheme to track attacker behaviors and forecast potential ransomware attacks. This solution implements malicious domain prediction (from observed patterns) and uses time series prediction, i.e., via \textit{hidden Markov models} (HMM) and \textit{autoregressive integrated moving averages} (ARIMA). However, only the Cerber ransomware family is analyzed here.  Meanwhile, \cite{subedi2018} uses data mining to match multi-level code components (extracted via reverse engineering) to identify unique rules. This static analysis approach is termed as CRSTATIC and operates on three code levels (assembly, library, and function). Specifically, the \textit{frequency patter} (FP) growth algorithm is also used, and families are differentiated using various attributes, e.g., propagation strategy, date appeared, cryptographic techniques, and C\&C servers.

Meanwhile, the \textit{digital forensic readiness} (DFR) solution in \cite{singh2018} defines mechanisms for secure communication, forensic soundness, and decryption. This scheme uncovers digital data evidence from computer hosts and networks and then extracts/stores this data in a secure database for offline analysis. Investigator reports are then generated on actions and processes performed by authorized entities. Also, \cite{zhu2020} assesses the \textit{VirusTotal} anti-malware-engine by analyzing 100 papers to gauge how this repository is used, i.e., for data pre-processing, label aggregation, engine independence, high-reputation engines, malware coverage, data sharing, etc. The authors conclude that most studies use a common threshold or a trusted set of vendors to classify a malicious file. Also, most efforts only take a single results screenshot and do not check for changes (leading to false positives). A two-year study in \cite{y.huang2018} also analyzes start-to-finish ransomware payment processing and concealment methods. Several families are tested on 4 platforms to obtain memory dumps and extract cryptocurrency addresses. Results show that hackers collected over \$16 million in ransom payments from almost 20,000 victims, with Bitcoin being the most common payment option. Also, \cite{laszka2017} presents an economic assessment of ransomware attacks using a game theoretic model with two groups (victims and hackers). The victim's decisions are analyzed (using real attacks) and the best solutions outlined. The authors conclude that paying ransoms encourages attacks and hence it is better to spend on data backups. Meanwhile, \cite{camelia2019} studies 1,180 American adults affected by ransomware attacks between 2016 and 2017. Of these, the average ransom demand was \$530, of which 4\% were paid. A self-assessment questionnaire is also used to develop a risk assessment model for users.  Finally, \cite{kharraz2015} analyzes ransomware attacks from 2006-2014 and notes that effective prevention is not necessarily complicated (despite the large number of families involved).  Specifically, mitigation techniques can prevent many attacks, e.g., such as API call monitoring, file system activity monitoring, decoy resources, etc. Such analysis of file system activities can also improve readiness against zero-day ransomware threats. 

\section{Malware Authorship Attribution}
\label{section_attribution}
Malware authorship attribution analyzes the key stylistic features of malware code to identify its authors (creators).  This information can help with digital forensics tasks or tracking down malactors. However, effective authorship attribution requires the availability of malware source code, not just binaries, and this can be a major challenge in the real world settings.  Indeed, many code features/styles may get obfuscated during the compilation process to generate binaries, and hackers themselves may take steps to remain hidden.  Some related works in this ransomware domain are now presented.

\subsection{Source Code Analysis}
A source code analysis scheme for multi-author identification, Multi-Xin, is presented in \cite{abuhamad2020}. This solution performs authorship verification, segment authorship identification, and authorship identification. Results show that Multi-X can identify coding styles using small code segments, as well as multiple authors in a single source code (with 86\% accuracy). Meanwhile \cite{aylin2015} uses RF classifiers to ``de-anonymize'' C/C++ programmers based on their coding styles. Namely, the work analyzes code derived from abstract syntax trees and yields accuracy rates of 94\% (1,600 authors) and 98\% (250 authors). However many programmers use similar techniques, and hence larger samples can lead to overfitting. Also, the authors in \cite{krsul1997} note that the likelihood of two programmers writing similar code for the same task is very low. Hence the study analyzes coding characteristics to distinguish between authors, e.g., via inline comments, blocked comments, space indentation, lower-case only, upper-case only, etc. Gaussian likelihood and NN-based methods are also used to identify authors, with error rates under 2\%.  Meanwhile \cite{spafford1993} uses software forensics to analyze code fragments and track identities. Different aspects are considered here, including language, formatting, special features, comment styles, variable names, scoping, execution paths, etc. However, a lot of malware code uses snippets from multiple authors, and programmers can also disguise their code to complicate attribution. 

\subsection{Binary Analysis}
\textit{Binary code analysis} (BCA) performs direct analysis of binary executables without 
access to the source code. Along these lines, \cite{xue2019} notes that BCA can be used for binary code clone detection, function recognition, malware detection, vulnerability discovery, and authorship recognition. However, since the raw extracted features cannot be directly used in supervised ML models, the authors further partition them to generate embedded vectors, i.e., graph- and code-based. Meanwhile, \cite{rosenblum2011} focuses on program authorship attribution and deciphers details using code characteristics. The objectives here include identifying program authors and finding stylistic similarities between programs written by unknown authors. A feature set is extracted (including N-grams, idioms, graphlets, super-graphlets, and library calls) and then used to train a SVM classifier. Results show that the scheme can successfully identify authors in over 10,000 samples. Another binary analysis study is also presented in \cite{meng2017}. Here the authors state that attribution can provide key information on malware forensics, software supply chain risk management, and software plagiarism detection. However, most techniques assume that a binary is written by a single author, which is generally not the case since most software (malware) is developed by a team. An empirical study is then presented using data from three large open-source projects. Specifically, the researchers develop a method to capture programming styles at the block level by looking at control and data flows. Blocks are then compared to determine if they are written by one or more authors.  This scheme achieves a block detection rate of 65\% for 284 authors.

\section{Ransomware Analysis Tools \& Facilities}
\label{tools}
Given the wide range of technical tasks involved in ransomware analysis, related studies have utilized many different software tools and facilities. These are briefly reviewed here:
\begin{itemize}
    \item \underline{\textbf{Malware Repositories}}: Many researchers have compiled their own independent ransomware datasets. However, others have used facilities such as \textit{VirusTotal} and \textit{VirusShare} which host large malware repositories with many ransomware families. These malware binaries are typically uploaded by users and can be downloaded for detailed analysis.
      \item \underline{\textbf{Raw Trace Capture}}: Sandbox and VM tools are widely used to analyze ransomware binaries and capture trace files.  Most notably, the \textit{Cuckoo} sandbox is a very popular option for Windows-based testing, and the \textit{Triage} facility (www.tri.ge) also hosts a powerful online sandbox. Specifically, researchers can use this latter resource to upload and run malware binaries and also download existing (pre-loaded/pre-processed) binaries and reports.
        \item \underline{\textbf{Pre-Processing/Feature Extraction}}: ML methods require extensive data pre-processing to select and extract training features.  Earlier, most researchers developed their own customized code for these purposes, e.g., using C/C++, Python, Java, etc. However, many ML packages (detailed next) already provide extensive features to support such processing. The Java-based open-source \textit{Pandas} toolkit also offers advanced data manipulation and transformation support for labeled datasets.
        \item \underline{\textbf{Machine Learning}}: A host of open source ML packages are now available.  These include toolkits such as \textit{TensorFlow}, \textit{Scikit}, \textit{PyTorch}, \textit{Weka}, and \textit{Keras.io}.  Collectively, these solutions provide full support for almost all types of (supervised, unsupervised) ML algorithms, i.e., such as linear regression, $k$-NN, $k$ means clustering, decision trees, RF, SVM, and most NN-based variants (baseline NN, CNN, DNN, recurrent NN, LSTM, Bi-LSTM, etc).
\end{itemize}
    
\section{Open Challenges}
\label{challenges}
Ransomware will continue to post a threat well into the future. Hence it is crucial to keep pace with these evolving scenarios and develop effective detection and classification frameworks. Overall, the works surveyed herein clearly represent a significant set of contributions in this space. However, there are still many open challenges that need to be addressed, and some of these are highlighted briefly.

Foremost, there is a pressing need to standardize ransomware datasets and testcase scenarios. Indeed, it is very difficult to compare existing schemes since most researchers have analyzed different subsets of ransomware using their own datasets. Hence it is important to identify a subset of the most relevant ransomware families and build/maintain a repository of binary downloads for each. Furthermore, associated testing and performance parameters also need to be specified, e.g., such as sandbox or VM run-times, evaluation metrics, etc.  These overall steps will help improve reproducibility and enable proper comparative analysis.

Additionally, it is vital to address emerging scalability and privacy concerns for ML-based ransomware detection and classification. Namely, even though many solutions have been proposed, their practicality in real-world settings has not been fully considered. For example, many users may be unwilling to share their detailed log files (traces) for external analysis. At the same time, complete local pre-processing of raw data may be too burdensome for local host processors.  Furthermore, it may become difficult to implement centralized data collection and ML computation at one site. Hence there is a further need to develop proper frameworks to address these concerns.

\bibliographystyle{IEEEtran}
\bibliography{references.bib}

\begin{thebibliography}{10}
\providecommand{\url}[1]{#1}
\csname url@samestyle\endcsname
\providecommand{\newblock}{\relax}
\providecommand{\bibinfo}[2]{#2}
\providecommand{\BIBentrySTDinterwordspacing}{\spaceskip=0pt\relax}
\providecommand{\BIBentryALTinterwordstretchfactor}{4}
\providecommand{\BIBentryALTinterwordspacing}{\spaceskip=\fontdimen2\font plus
\BIBentryALTinterwordstretchfactor\fontdimen3\font minus
  \fontdimen4\font\relax}
\providecommand{\BIBforeignlanguage}[2]{{%
\expandafter\ifx\csname l@#1\endcsname\relax
\typeout{** WARNING: IEEEtran.bst: No hyphenation pattern has been}%
\typeout{** loaded for the language `#1'. Using the pattern for}%
\typeout{** the default language instead.}%
\else
\language=\csname l@#1\endcsname
\fi
#2}}
\providecommand{\BIBdecl}{\relax}
\BIBdecl

\bibitem{cisco2016}
NA, ``Ransomware defense validated design guide,'' \emph{Cisco Systems}, 2016.

\bibitem{trends2022}
------, ``Ransomware facts, trends \& statistics for 2022,'' \emph{Safety
  Detectives}, 2022.

\bibitem{kapoor2022}
A.~Kapoor, ``Ransomware detection, avoidance, and mitigation scheme: A review
  and future directions,'' \emph{Sustainability}, vol.~14, no.~1, Dec. 2021.

\bibitem{law2021}
NA, ``Senate bill to mandate cyberattack, ransomware payment reporting,''
  \emph{Bloomberg Government}, September 28, 2021.

\bibitem{berrueta2019}
E.~Berrueta, D.~Morato, E.~Magaña, and M.~Izal, ``A survey on detection
  techniques for cryptographic ransomware,'' \emph{IEEE Access}, vol.~7, pp.
  144\,925--144\,944, October 2019.

\bibitem{mouss2021}
R.~Moussaileb, N.~Cuppens, J.-L. Lanet, and Bouder, ``A survey on windows-based
  ransomware taxonomy and detection mechanisms,'' \emph{ACM Computing Surveys},
  vol.~54, no.~6, July 2022.

\bibitem{almash2019}
A.~Almashhadani, M.~Kaiiali, S.~Sezer, and P.~O’Kane, ``A multi-classifier
  network-based crypto ransomware detection system: A case study of locky
  ransomware,'' \emph{IEEE Access}, vol.~7, no.~1, pp. 47\,053--47\,067, 2019.

\bibitem{mulders2017}
D.~Mulders, ``Network based ransomware detection on the samba protocol,''
  \emph{MS Thesis, Dept. of Mathematics, TU Eindhoven}, 2017.

\bibitem{morato2018}
D.~Morato, E.~Berrueta, E.~Magaña, and M.~Izal, ``Ransomware early detection
  by the analysis of file sharing traffic.'' \emph{Journal of Network and
  Computer Applications}, vol. 124, no.~1, pp. 14--32, 2018.

\bibitem{cusak2018}
G.~Cusack, O.~Michel, and E.~Keller, ``Machine learning-based detection of
  ransomware using sdn,'' in \emph{SDN-NFV 2018}, Tempe, AZ, March 2018.

\bibitem{lokuketagoda2018}
B.~Lokuketagoda, M.~Weerakoon, U.~Kuruppu, A.~Senarathne, and K.~Abeywardena,
  ``R-killer: An email based ransomware protection tool,'' in \emph{ICCSE
  2018}, Singapore, July 2018.

\bibitem{homayoun2019}
S.~Homayoun, A.~Dehghantanha, M.~Ahmadzadeh, and S.~Hashemi, ``Drthis: Deep
  ransomware threat hunting and intelligence system at the fog layer,''
  \emph{Future Generation Computer Systems}, pp. 94--104, Jan. 2019.

\bibitem{kurniawan2018}
K.~Ade and R.~Imam, ``Detection and analysis cerber ransomware based on network
  forensics behavior,'' \emph{International Journal of Network Security},
  vol.~20, no.~5, 2018.

\bibitem{su2019}
D.~Su, J.~Liu, X.~Wang, W.~Wang, and P.~O’Kane, ``Detecting android
  locker-ransomware on chinese social networks,'' \emph{IEEE Access}, vol.~7,
  pp. 20\,381--20\,393, 2019.

\bibitem{kao2018}
Kao and S.~Hsiao, ``The dynamic analysis of wannacry ransomware,'' in
  \emph{ICACT 2018}, South Korea, February 2018.

\bibitem{roy2021}
K.~C. Roy and Q.~Chen, ``Deepran: Attention-based bilstm and crf for ransomware
  early detection and classification. information systems frontiers,''
  \emph{Information Systems Frontiers}, vol.~0, pp. 1--17, 2021.

\bibitem{krzysztof2018}
K.~Cabaj, M.~Gregorczyk, and W.~Mazurczyk, ``Software-defined networking-based
  crypto ransomware detection using http traffic characteristics,''
  \emph{Computers in Electrical Eng.}, vol.~66, pp. 353--368, 2019.

\bibitem{alhawi2018}
O.~Alhawi, J.~Baldwin, and A.~Dehghantanha, ``Leveraging machine learning
  techniques for windows ransomware network traffic detection,'' \emph{Advances
  in Information Security}, p. 93–106, July 2018.

\bibitem{kharraz2016}
A.~Kharaz, S.~Arshad, C.~Mulliner, W.~Robertson, and E.~Kirda, ``{UNVEIL}: A
  {Large-Scale}, automated approach to detecting ransomware,'' in \emph{USENIX
  Security 2016}, Austin, TX, August 2016.

\bibitem{kolodenker2017}
K.~Eugene, K.~Wil, S.~Gianluca, and E.~Manuel, ``Paybreak: Defense against
  cryptographic ransomware,'' in \emph{Asia CCS 2017}, Abu Dhabi, UAE, April
  2017.

\bibitem{continella2016}
A.~Continella, A.~Guagnelli, G.~Zingaro, G.~Pasquale, A.~Barenghi, S.~Zanero,
  and F.~Maggi, ``Shieldfs: a self-healing, ransomware-aware filesystem,'' in
  \emph{Annual Computer Security Applications Conference (ACSAC) 2016}, Los
  Angeles, CA, December 2016.

\bibitem{kharraz2017}
A.~Kharraz and E.~Kirda, ``Redemption: Real-time protection against ransomware
  at end-hosts,'' in \emph{RAID 2017}, Atlanta, GA, October 2017.

\bibitem{baek2017}
S.~Baek, Y.~Jung, D.~Mohaisen, S.~Lee, and D.~Nyang, ``Ssd-insider: Internal
  defense of solid-state drive against ransomware with perfect data
  recoveryquestio,'' in \emph{ICDCS 2018}, Vienna, Austria, July 2018.

\bibitem{sinha2018}
S.~Sinha, M.~Alam, S.~Bhattacharya, D.~Mukhopadhyay, A.~Chattopadhyay, and
  S.~Dutta, ``Rapper: Ransomware prevention via performance counters,'' in
  \emph{Kangacrypt 2018}, Adelaide, Australia, December 2018.

\bibitem{molina2021}
R.~M.~A. Molina, S.~Torabi, K.~Sarieddine, E.~Bou-Harb, N.~Bouguila, and
  C.~Assi, ``On ransomware family attribution using pre-attack paranoia
  activities,'' \emph{IEEE Transactions on Network and Service Management},
  vol.~19, no.~1, pp. 19--36, 2022.

\bibitem{alsabeh2020}
A.~AlSabeh, H.~Safa, E.~Bou-Harb, and J.~Crichigno, ``Exploiting ransomware
  paranoia for execution prevention,'' in \emph{IEEE ICC 2020}, Dublin,
  Ireland, June 2020.

\bibitem{molina2022}
R.~M.~A. Molina, ``Rpm: Ransomware prevention and mitigation using operating
  systems sensing tactics,'' \emph{submitted}, 2022.

\bibitem{huang2017}
J.~Huang, J.~Xu, X.~Xing, P.~Liu, and M.~Qureshi, ``Leveraging intrinsic flash
  properties to defend against encryption ransomware,'' in \emph{ACM SIGSAC
  2017}, Dallas, Texas USA, October 2017.

\bibitem{poudyal2018}
S.~Poudyal, K.~P. Subedi, and D.~Dasgupta, ``A framework for analyzing
  ransomware using machine learning.'' \emph{IEEE 2018 SSCI}, Nov. 2018.

\bibitem{zhang2019}
H.~Zhang, X.~Xiao, F.~Mercaldo, S.~Ni, F.~Martinelli, and A.~Sangaiah,
  ``Classification of ransomware families with machine learning based onn-gram
  of opcodes.'' \emph{Future Generation Computer Systems}, vol.~90, pp.
  211--221, 2019.

\bibitem{ming2017}
J.~Ming, D.~Xu, Y.~Jiang, D.~Wu, Debdeep, A.~Chattopadhyay, and S.~Dutta,
  ``Binsim: Trace-based semantic binary diffing via system call sliced segment
  equivalence checking,'' in \emph{USENIX Security Symposium 2017}, Vancouver,
  Canada, August 2017.

\bibitem{chen2017}
Z.~Chen, H.~Kang, S.~Yin, and S.~Kim, ``Automatic ransomware detection and
  analysis based on dynamic api calls ﬂow graph,'' in \emph{RACS 2019},
  Chongqing, China, September 2019.

\bibitem{quinkert2018}
F.~Quinkert, T.~Holz, K.~S. M.~T. Hossain, E.~Ferrara, and K.~Lerman, ``Raptor:
  Ransomware attack predictor.'' in \emph{arXiv 1803.01598}, Mar. 2018.

\bibitem{subedi2018}
K.~Subedi, D.~R. Budhathoki, and D.~Dasgupta, ``Forensic analysis of ransomware
  families using static and dynamic analysis,'' in \emph{SADFE 2018}, San
  Francisco, CA, USA, May 2018.

\bibitem{singh2018}
A.~Singh, I.~Adeyemi, and H.~Venter, ``Digital forensic readiness framework,''
  in \emph{ICDF2C 2018}, New Orleans, LA, USA, September 2018.

\bibitem{zhu2020}
S.~Zhu, J.~Shi, L.~Yang, B.~Qin, Z.~Zhang, L.~Song, and G.~Wang, ``Measuring
  and modeling the label dynamics of online anti-malware engines,'' in
  \emph{USENIX Security 2020}, Virtual, Online, August 2020.

\bibitem{y.huang2018}
D.~Y. Huang, M.~M. Aliapoulios, V.~G. Li, L.~Invernizzi, E.~Bursztein,
  K.~McRoberts, J.~Levin, K.~Levchenko, A.~C. Snoeren, and D.~McCoy, ``Tracking
  ransomware end-to-end,'' in \emph{IEEE Symposium on Security and Privacy (SP
  2018)}, 2018, pp. 618--631.

\bibitem{laszka2017}
A.~Laszka, S.~Farhang, and J.~Grossklags, ``On the economics of ransomware,''
  in \emph{GameSec 2017}, Vienna, Austria, July 2017.

\bibitem{camelia2019}
C.~Simoiu, C.~Gates, J.~Bonneau, and S.~Goel, ``"i was told to buy a software
  or lose my computer. i ignored it": A study of ransomware,'' in \emph{SOUPS
  2019}, Santa Clara, CA, August 2019.

\bibitem{kharraz2015}
A.~Kharraz, W.~Robertson, D.~Balzarotti, L.~Bilge, and E.~Kirda, ``Cutting the
  gordian knot: A look under the hood of ransomware attacks,'' in \emph{DIMVA
  2015}, Milan, Italy, July 2015.

\bibitem{abuhamad2020}
M.~Abuhamad, T.~Abuhmed, D.~Nyang, and D.~Mohaisen, ``Identifying multiple
  authors from source code files,'' \emph{Proceedings on Privacy Enhancing
  Technologies}, pp. 25--41, July 2020.

\bibitem{aylin2015}
A.~Caliskan-Islam, R.~Harang, A.~Liu, A.~Narayanan, C.~Voss, F.~Yamaguchi, and
  R.~Greenstadt, ``De-anonymizing programmers via code stylometry,'' in
  \emph{USENIX Security 2015}, Austin, TX, August 2015.

\bibitem{krsul1997}
I.~Krsul and E.~H. Spafford, ``Authorship analysis: identifying the author of a
  program,'' \emph{Computers and Security}, vol.~16, no.~3, pp. 233--257, 1997.

\bibitem{spafford1993}
E.~H. Spafford and S.~A. Weeber, ``Software forensics: Can we track code to its
  authors?'' \emph{Computers and Security}, vol.~12, no.~6, pp. 585--595, 1993.

\bibitem{xue2019}
H.~Xue, S.~Sun, G.~Venkataramani, and T.~Lan, ``Machine learning-based analysis
  of program binaries: A comprehensive study,'' \emph{IEEE Access}, vol.~7, pp.
  65\,889--65\,912, 2019.

\bibitem{rosenblum2011}
N.~E. Rosenblum, X.~Zhu, and B.~P. Miller, ``Who wrote this code? identifying
  the authors of program binaries,'' in \emph{ESORICS 2011}, Leuven, Belgium,
  September 2011.

\bibitem{meng2017}
X.~Meng, B.~P. Miller, and K.~Jun, ``Identifying multiple authors in a binary
  program,'' in \emph{ESORICS 2017}, Oslo, Norway, September 2017.

\end{thebibliography}

\end{document}